\documentclass[sigconf, nonacm,anonymous=false]{acmart}

\newcommand\vldbdoi{10.14778/3685800.3685900}
\newcommand\vldbpages{4461 - 4464}
\newcommand\vldbvolume{17}
\newcommand\vldbissue{12}
\newcommand\vldbyear{2024}
\newcommand\vldbauthors{\authors}
\newcommand\vldbtitle{\shorttitle} 
\newcommand\vldbavailabilityurl{https://github.com/jls2007/OFL-W3}
\newcommand\vldbpagestyle{empty}

\usepackage{color}
\usepackage{graphicx}
\usepackage{import}
\usepackage{algorithm}
\usepackage{algpseudocodex}
\usepackage{tikz}
\usepackage{xparse}
\usepackage{subcaption}
\usepackage{multirow}
\usepackage{makecell}
\usepackage{bbding}
\usepackage{threeparttable}
\usepackage{listings}
\usepackage{pythonhighlight}
\usepackage{lipsum}
\usepackage{listings}
\usepackage{wrapfig}

\usepackage{xcolor}

\definecolor{codegreen}{rgb}{0,0.6,0}
\definecolor{codegray}{rgb}{0.5,0.5,0.5}
\definecolor{codepurple}{rgb}{0.58,0,0.82}
\definecolor{backcolour}{rgb}{0.95,0.95,0.92}

\lstdefinestyle{mystyle}{
	backgroundcolor=\color{backcolour},   
	commentstyle=\color{codegreen},
	keywordstyle=\color{magenta},
	numberstyle=\tiny\color{codegray},
	stringstyle=\color{codepurple},
	basicstyle=\ttfamily\footnotesize,
	breakatwhitespace=false,         
	breaklines=true,                 
	captionpos=b,                    
	keepspaces=true,                 
	numbers=left,                    
	numbersep=5pt,                  
	showspaces=false,                
	showstringspaces=false,
	showtabs=false,                  
	tabsize=2
}

\begin{document}
\title{OFL-W3: A One-shot Federated Learning System on Web 3.0}

\author{Linshan Jiang}
\email{linshan@nus.edu.sg}
\orcid{0000}
\affiliation{%
	\institution{National University of Singapore}
	\streetaddress{13 Computing Drive}
	%\city{Singapore}
	%\country{Singapore}
	\postcode{117417}
}

\author{Moming Duan}
\email{moming@nus.edu.sg}
\orcid{0000}
\affiliation{%
	\institution{National University of Singapore}
	\streetaddress{13 Computing Drive}
	%\city{Singapore}
	%\country{Singapore}
	\postcode{117417}
}

\author{Bingsheng He}
\email{hebs@comp.nus.edu.sg}
\orcid{0000}
\affiliation{%
	\institution{National University of Singapore}
	\streetaddress{13 Computing Drive}
	%\city{Singapore}
	%\country{Singapore}
	\postcode{117417}
}

\author{Yulin Sun}
\email{rain-forest@sjtu.edu.cn}
\orcid{0000}
\affiliation{%
	\institution{Shanghai Jiao Tong University}
	\streetaddress{13 Computing Drive}
	%\city{Singapore}
	%\country{Singapore}
	\postcode{117417}
}
\author{Peishen	Yan}
\email{peishenyan@sjtu.edu.cn}
\orcid{0000}
\affiliation{%
	\institution{Shanghai Jiao Tong University}
	\streetaddress{13 Computing Drive}
	%\city{Singapore}
	%\country{Singapore}
	\postcode{117417}
}
\author{Yang	Hua}
\email{Y.Hua@qub.ac.uk}
\orcid{0000}
\affiliation{%
	\institution{Queen's University Belfast}
	\streetaddress{13 Computing Drive}
	%\city{Singapore}
	%\country{Singapore}
	\postcode{117417}
}
\author{	Tao	Song}
\email{songt333@sjtu.edu.cn}
\orcid{0000}
\affiliation{%
	\institution{Shanghai Jiao Tong University}
	\streetaddress{13 Computing Drive}
	%\city{Singapore}
	%\country{Singapore}
	\postcode{117417}
}

%%
%% The "author" command and its associated commands are used to define the authors and their affiliations.
%\authoranon{\author{Ben Trovato}
%	\affiliation{%
%		\institution{Institute for Clarity in Documentation}
%		\streetaddress{P.O. Box 1212}
%		\city{Dublin}
%		\state{Ireland}
%		\postcode{43017-6221}
%	}
%	\email{trovato@corporation.com}}
%

%%
%% The abstract is a short summary of the work to be presented in the
%% article.
\begin{abstract}
Federated Learning (FL) addresses the challenges posed by data silos, which arise from privacy, security regulations, and ownership concerns. Despite these barriers, FL enables these isolated data repositories to participate in collaborative learning without compromising privacy or security. Concurrently, the advancement of blockchain technology and decentralized applications (DApps) within Web 3.0 heralds a new era of transformative possibilities in web development. As such, incorporating FL into Web 3.0 paves the path for overcoming the limitations of data silos through collaborative learning. However, given the transaction speed constraints of core blockchains such as Ethereum (ETH) and the latency in smart contracts, employing one-shot FL, which minimizes client-server interactions in traditional FL to a single exchange, is considered more apt for Web 3.0 environments. This paper presents a practical one-shot FL system for Web 3.0, termed OFL-W3. OFL-W3 capitalizes on blockchain technology by utilizing smart contracts for managing transactions. Meanwhile,  OFL-W3 utilizes the Inter-Planetary File System (IPFS) coupled with Flask communication, to facilitate backend server operations to use existing one-shot FL algorithms. With the integration of the incentive mechanism, OFL-W3 showcases an effective implementation of one-shot FL on Web 3.0, offering valuable insights and future directions for AI combined with Web 3.0 studies.
	
%In the Web 3.0 era, characterized by decentralization, user autonomy, and a user-centric approach, the integration of blockchain technology and decentralized applications (DApps) has unlocked significant transformative potential in web development. Federated Learning (FL), aligning with Web 3.0’s key features, presents as an innovative DApp with potential to revolutionize the next-generation Internet. Given the transaction speed limitations of backbone blockchains like Ethereum (ETH) and smart contract latency, one-shot FL, which limits client-server communication in conventional FL to a single round, is deemed more suitable for Web 3.0 applications. In this demo paper, we introduce a practical one-shot FL system on Web 3.0, named OFL-W3. In OFL-W3, we integrate blockchain technology employing smart contracts for transaction management and leveraging the Inter-Planetary File System (IPFS) for model sharing with Flask communication to support backend server executing existing one-shot FL algorithms.  With several different incentive mechanisms provided, OFL-W3 demonstrates practical one-shot FL implementation in Web 3.0, and providing insights and directions for AI+Web 3.0 research.
\end{abstract}

\maketitle

%%% do not modify the following VLDB block %%
%%% VLDB block start %%%

%%% do not modify the following VLDB block %%
%%% VLDB block start %%%
\pagestyle{\vldbpagestyle}
\begingroup\small\noindent\raggedright\textbf{PVLDB Reference Format:}\\
\vldbauthors. \vldbtitle. PVLDB, \vldbvolume(\vldbissue): \vldbpages, \vldbyear.\\
\href{https://doi.org/\vldbdoi}{doi:\vldbdoi}
\endgroup
\begingroup
\renewcommand\thefootnote{}\footnote{\noindent
	This work is licensed under the Creative Commons BY-NC-ND 4.0 International License. Visit \url{https://creativecommons.org/licenses/by-nc-nd/4.0/} to view a copy of this license. For any use beyond those covered by this license, obtain permission by emailing \href{mailto:info@vldb.org}{info@vldb.org}. Copyright is held by the owner/author(s). Publication rights licensed to the VLDB Endowment. \\
	\raggedright Proceedings of the VLDB Endowment, Vol. \vldbvolume, No. \vldbissue\ %
	ISSN 2150-8097. \\
	\href{https://doi.org/\vldbdoi}{doi:\vldbdoi} \\
}\addtocounter{footnote}{-1}\endgroup
%%% VLDB block end %%%

%\begingroup
%\renewcommand\thefootnote{}\footnote{\noindent
%This work is licensed under the Creative Commons BY-NC-ND 4.0 International License. Visit \url{https://creativecommons.org/licenses/by-nc-nd/4.0/} to view a copy of this license. For any use beyond those covered by this license, obtain permission by emailing \href{mailto:info@vldb.org}{info@vldb.org}. Copyright is held by the owner/author(s). Publication rights licensed to the VLDB Endowment. \\
%\raggedright Proceedings of the VLDB Endowment, Vol. \vldbvolume, No. \vldbissue\ %
%ISSN 2150-8097. \\
%\href{https://doi.org/\vldbdoi}{doi:\vldbdoi} \\
%}\addtocounter{footnote}{-1}\endgroup
%%%% VLDB block end %%%
%
%%%% do not modify the following VLDB block %%
%%%% VLDB block start %%%
%\ifdefempty{\vldbavailabilityurl}{}{
%\vspace{.3cm}
%\begingroup\small\noindent\raggedright\textbf{PVLDB Artifact Availability:}\\
%The source code, data, and/or other artifacts have been made available at \url{\vldbavailabilityurl}.
%\endgroup
%}
%%% VLDB block end %%%

\section{Introduction}

Federated Learning (FL)~\cite{pmlr-v54-mcmahan17a} marks a pioneering shift in machine learning, enabling collaborative model training directly within data silos. This innovative approach allows for the collaborative training of models across various data silos, safeguarding the privacy of local data while simultaneously building robust global models. Concurrently, the emergence of Web 3.0 revolutionizes our digital interactions and online value exchange mechanisms. Powered by blockchain technology and decentralized applications (DApps), Web 3.0 introduces significant breakthroughs in distributed video platforms and cloud storage services. Therefore, merging FL with Web 3.0 introduces novel pathways for data silos to practically engage in the collaborative machine learning process with incentives.

%The advent of Web 3.0 has fundamentally transformed our digital interactions and the ways in which value is exchanged online. Anchored by blockchain technology and decentralized applications (DApps), Web 3.0 has pioneered advancements in distributed video platforms and cloud storage. It has notably flourished in transactional spheres such as Decentralized Finance (DeFi) and Digital Collectibles, including NFTs. To further energize Web 3.0 and drive its mass adoption, there is a pressing need for applications that address real-world needs.

%Simultaneously, Federated Learning (FL) represents a trailblazing approach in machine learning, by collaborative model training at the data source. This innovation facilitates collective model training across numerous distributed devices, ensuring the privacy of local data while constructing robust models globally. In several respects, FL embodies the principles of Web 3.0; it empowers users and organizations to tap into collective intelligence without relinquishing personal data ownership. The outcome depends the contributions of each participant, ultimately yielding benefits for all involved with incentives.

Standard FL algorithm FedAvg~\cite{pmlr-v54-mcmahan17a} requires a multitude of communication rounds for effective global model training, leading to considerable communication overhead, increased privacy risks, and a greater demand for fault tolerance. One-shot FL approaches ~\cite{yurochkin2019bayesian,guha2019one,diao2023towards}, which streamline client-server communication into a solitary round, offer a promising yet complex solution to mitigate these challenges with a tolerable impact on global model quality. Additionally, within the context of Web 3.0 applications, the transaction speed limitations of contemporary commercial blockchains such as Ethereum (ETH)~\cite{wood2014ethereum}, coupled with the high transaction costs (e.g., gas fees) on Web 3.0, render one-shot FL a viable option.

The practical implementation of one-shot FL on Web 3.0 encounters two significant challenges. Firstly, given that Web 3.0 research is still in its nascent stages, the fusion of Web 3.0 and FL, including the functionality and roles within this integration, remains an ambiguous issue. Secondly, considering the substantial gas fees~\cite{carl2020ethereum}  associated with transactions on ETH, it necessitates the simplification of smart contract designs. In other words, complex operations and the storage of models within smart contracts should be minimized to manage costs effectively.

To solve these challenges, in this demonstration, we present OFL-W3, a novel one-shot Federated Learning (FL) system optimized for Web 3.0. OFL-W3 categorizes data silos into two roles: model buyers, who lead the one-shot FL process and supply tokens for robust models, and model owners, who use their private data to contribute models to the one-shot FL process in exchange for tokens. To address storage and smart contract complexity challenges on Web 3.0, we leverage the Inter-Planetary File System (IPFS)~\cite{benet2014ipfs} for efficient model sharing. Furthermore, we employ PFNM~\cite{yurochkin2019bayesian} as the one-shot FL algorithm and Leave-one-out as the incentive mechanism for illustration. To showcase our system, OFL-W3 includes a distributed application (DApp) built with React for the front end and Flask for backend services, integrated with the Google Chrome browser and MetaMask wallet extension. This configuration enables model owners to participate in the FL system and receive token rewards with no prior blockchain or Web 3.0 knowledge, while model buyers can access decentralized models to build robust global models, while maintaining data privacy and security. Our contribution can be summarized as follows.

 Our system offers a user-friendly DApp that allows data silos to participate in the one-shot FL learning system, either as model buyers or model owners. Designed for simplicity and ease of use, OFL-W3 enables anyone, regardless of their knowledge of blockchain or Web 3.0, to share their models or obtain high-quality ML models.

 %Through experiments and demonstrations, we showcase our system's ability to effectively utilize decentralized shared models on the data silos, delivering high-quality models to the model buyers. Our experimental results offer valuable insights for future research in the convergence of AI and Web 3.0.

%The rest of this paper is organized as follows. Section 2 introduces
%related works. Section 3 presents an overview of OFL-W3.
%Section 4 demonstrates the experiments on OFL-W3. Section 5 concludes our paper.
\section{Related Works}
\textbf{One-shot Federated Learning.} One-shot Federated Learning (FL) represents a cutting-edge and promising avenue of research, distinguished by its notably low communication cost. The initial exploration into one-shot FL~\cite{guha2019one} presents a method that aggregates local models into an ensemble to formulate the final global model, followed by the application of knowledge distillation utilizing public data.  Researchers introduce PFNM~\cite{yurochkin2019bayesian}, a Bayesian probabilistic framework specifically tailored for multi-layer perceptrons.
% A significant advancement using Knowledge Distillation is made by DENSE~\cite{zhang2022dense}, which introduces the use of a generator to yield synthetic datasets server-side, eliminating the requirement for public datasets in the distillation phase. 
Lastly, FedOV~\citep{diao2023towards} ventures into tackling cases of label skew, marking another step forward in the evolution of one-shot FL approaches. %Based on the code availability and the simplicity of one-shot federated algorithms, we adopt PFNM~\cite{yurochkin2019bayesian} in our demonstration. 

\noindent \textbf{Blockchain-enabled Federated Learning.} Blockchain-enabled Federated Learning (FL) has conventionally addressed the privacy and security challenges inherent in FL frameworks. For instance, Blockchain-based PPFL~\cite{awan2019poster} leverages blockchain technology to trace models and prevents tampering by unauthorized individuals. BlockFlow~\cite{mugunthan2020blockflow} addresses concerns related to dishonest participants by employing blockchain and consensus mechanisms.
% for decentralized contribution assessment.

% Similarly, BlockFLA~\cite{desai2021blockfla} and OrderlessFL~\cite{nasirifard2022orderlessfl} utilize cryptographic hashes maintained on the public chain to ensure the integrity of models.
These approaches to FL for Web 3.0 have certain limitations, including their dependency on local blockchains, lack of public code, which hinders system evaluation, and potentially inaccurate estimated gas fees for contemporary commercial blockchains on Web 3.0. The absence of DApps restricts engagement to Web 3.0 specialists. Additionally, sharing models directly on the blockchain, as seen in several studies, increases numerical execution costs on smart contracts, challenging their widespread adoption. Moreover, relying on traditional FL algorithms introduces substantial overhead from multi-round communication over the blockchain.
%\subsection{Web3 applications}

\section{System}
%This section provides an overview of OFL-W3, highlighting the roles of two key entities and detailing the workflows for OFL-W3 step-by-step.

\subsection{System Overview}

\begin{figure}
	\centering
	\vspace{-1em}
	\includegraphics[width=0.88\linewidth]{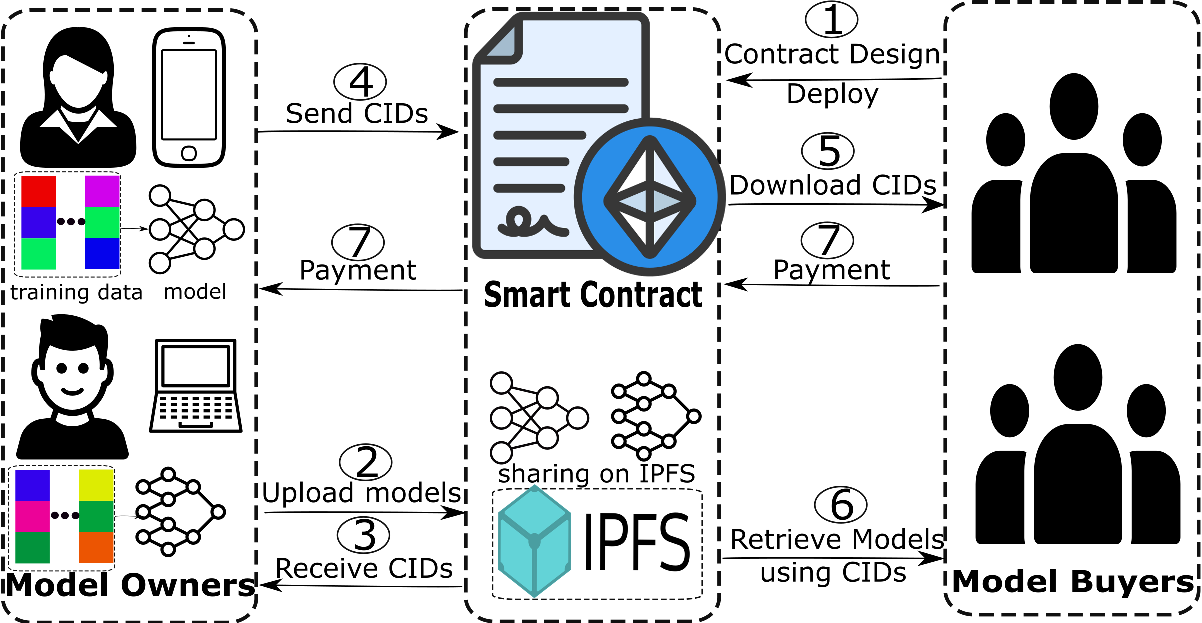}
	\caption{The System Overview of OFL-W3.}
	\vspace{-1em}
	\label{fig:overview}
	\vspace{-1em}
\end{figure}
As shown in Fig.~\ref{fig:overview}, OFL-W3 consists of the following
two entities.
\noindent \textbf{Model Buyers.} Model buyers have  demands for high-quality ML
	models. They aggregate the shared models on a one-shot FL algorithm through OFL-W3 to improve model quality. 
	
\noindent \textbf{Model Owners.} Model owners can participate in model aggregation via OFL-W3, which requires sufficient incentives. Note that the models may come from the local training if the model owner also performs as the data owner, or fine-tuned/transferred from existing backbone models on their own techniques.  

In our system, model buyers benefit from improved model quality via the one-shot FL paradigm, at the cost of spending digital tokens, including transaction fees or gas fees. Model owners gain by acquiring tokens, but face costs from training models with private data or adapting existing models, in addition to gas fees.

\begin{figure}[htpb]
	\centering
	\vspace{-1em}
	\begin{lstlisting}
		pragma solidity ^0.8.7;
		contract CidStorage {
			uint256 public cidCount;
			...
			function uploadCid(string memory cid) public {
				cids[cidCount] = cid;
				cidCount++;
				emit CidUploaded(cid);}
			...
			function getCid(uint256 index) public view returns (string memory) {
				require(index < cidCount, "Invalid CID index");
				return cids[index];
			}
			...
		}
	\end{lstlisting}
	\vspace{-1em}
	\caption{The partial example solidity codes of smart contract.}
	\label{fig:smart_contract}
	\vspace{-1pt}
\end{figure}

\begin{figure}[htbp]
	\centering
	\vspace{-1em}
	% Subfigure A
	\begin{subfigure}[b]{0.35\linewidth}
		\centering
		\includegraphics[width=0.7\linewidth]{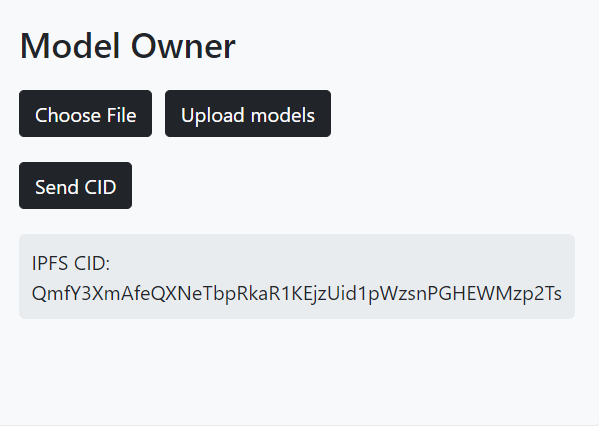}
		\caption{Model Owners}
		\label{fig:owner_interface}
	\end{subfigure}
	%\hfill % This will add a little bit of space between the subfigures horizontally.
	% Subfigure B8
	\begin{subfigure}[b]{0.55\linewidth}
		\centering
		\includegraphics[width=0.8\linewidth]{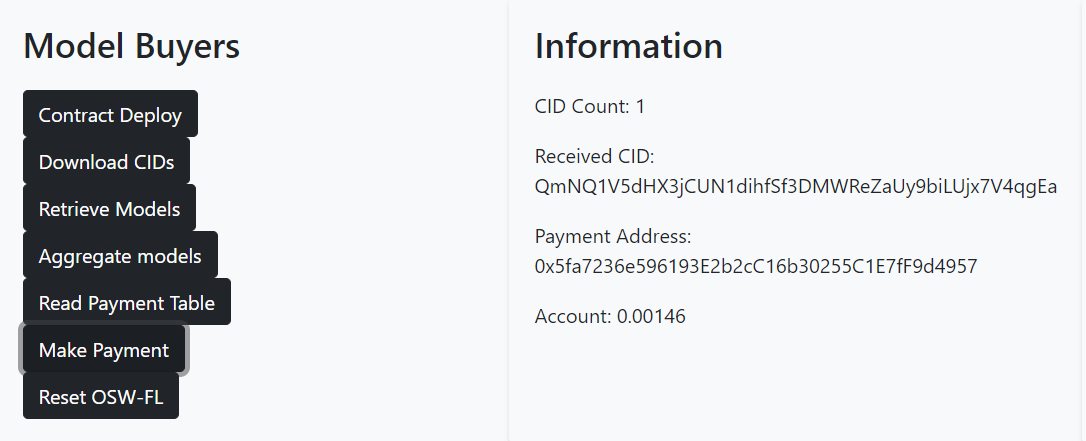}
		\caption{Model Buyers}
		\label{fig:buyer_interface}
	\end{subfigure}
	\vspace{-1em}
	\caption{Interfaces in OFL-W3.}
	\label{fig:whole_interface}
	\vspace{-1em}
\end{figure}

\subsection{Workflow}

%Now, we detail the concrete steps and their realizations.

\noindent \textbf{Step 1. Contract Design and Deploy.} Model buyers design and deploy a smart contract tailored to a specific one-shot FL algorithm on the Sepolia ETH test network\footnotemark, specifying ML tasks, model structures, initial models, and necessary auxiliary information if the one-shot FL algorithm requires. They outline the payment in tokens and launch the contract on a commercial blockchain. 
%In this paper, we hereby demonstrate the smart contract under the standard of ERC-20 on the Sepolia ETH test network\footnotemark, with the contract address made public. Partial example codes are depicted in Figure~\ref{fig:smart_contract}.   

\footnotetext{Note that the deployed contract can be directly transferred on the ETH mainnet since they use the same standard. However, due to the high price of ETH, we mainly show our system on SepoliaETH, one of ETH testnet. Now, 1 Sepolia ETH is around $\$0.00006874$ while 1 ETH is around $\$3,466$.} %In the rest of the paper, we use ETH instead of SepoliaETH. }

\noindent \textbf{Step 2. Upload Models.} Model owners find the smart contract using its address, agree to participate in the one-shot FL system, and prepare models according to the contract's specifications, including any necessary auxiliary information. They then upload these prepared models to the IPFS, with or without additional data.

\noindent \textbf{Step 3. Receive CIDs.} In the IPFS, a distributed file system, models are assigned 32-byte Content Identifiers (CIDs) through cryptographic hashes. This system ensures the unique accessibility and integrity of uploaded content, allowing for efficient model retrieval.

\noindent \textbf{Step 4. Send CIDs.} After receiving CIDs from IPFS, model owners submit these identifiers to the blockchain through the smart contract. This method conserves on-chain space, with each model occupying only 256 bits. As a comparison, at least Kb-level storage is needed if directly saving the model on the blockchain~\cite{awan2019poster,desai2021blockfla}, which proves to be impractical within the ETH network.

\noindent \textbf{Step 5. Download CIDs.} Model owners download the CIDs of all models shared via the smart contract, involving a process free of gas fees since it makes no data modification on the blockchain.

\noindent \textbf{Step 6. Retrieve Models.} After receiving the CIDs, the model owners can retrieve models with/without any auxiliary information. The retrieved models are used for the one-shot FL algorithm.

\noindent \textbf{Step 7. Payment.} The model buyers aggregate the retrieved model using its own one-shot FL algorithm, as denoted in the smart contract. The model buyers can adopt their own backend workstation/server to accelerate one-shot FL algorithm by using Flask to interact with the backend workstation. In this demonstration, we adopt PFNM~\cite{yurochkin2019bayesian} to aggregate the models. Then it assesses each participant's marginal contribution,  like Leave-one-out (LOO), to pay the calculated tokens. 
%PFNM employs a Beta-Bernoulli process to aggregate local models, quantifying the degree of alignment between global and local parameters.  
%Following aggregation, model buyers achieve a high-quality model and calculate individual payments for model owners using incentive payment functions, like Leave-one-out (LOO), which assesses each participant's marginal contribution. 
%Subsequently, the calculated token amounts are distributed to model owners through the smart contract. 

For the Dapp, the buyer's interface including Step 2 and Step 4 is illustrated in Fig.~\ref{fig:buyer_interface},  while the owner's interface including Step 1,2,5,6 and 7  is illustrated in Fig.~\ref{fig:owner_interface}. The simplicity of the interface  enables anyone with/without any knowledge of blockchain or Web 3.0 to use OFL-W3 by clicking buttons.

\section{Demonstrations and Experiments}
In our demo, we simulate a scenario with ten model owners and a model buyer using a server with two NVIDIA RTX A5000 GPUs to run the PFNM one-shot FL algorithm, targeting to develop a high-quality model with a total cost of 0.01 ETH (approximately \$34). The experiment utilizes the MNIST dataset and a neural network with three multi-layer perceptron layers (784, 100, 10). To mimic realistic non-IID data distributions, we use the data partitioning techniques in PFNM~\cite{yurochkin2019bayesian}. The local model training settings include a batch size of 64, a learning rate of 0.001, and 10 local epochs.

\subsection{Model Performance}
\begin{figure}
	\centering
	\vspace{-1em}
	\includegraphics[width=0.9\linewidth]{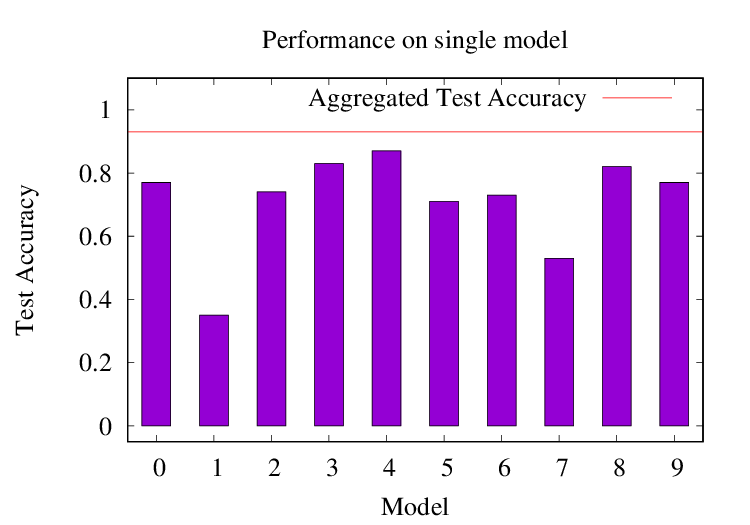}
	\vspace{-1em}
	\caption{Single local model quality among 10 model owners.}
	\label{fig:table}

\end{figure} 

Figure~\ref{fig:table} presents the quality of local models as evaluated by their test performance. This highlights the issue where, if a model owner is unable to effectively aggregate models from all participants, an individually trained model suffers from inadequate training data, leading to suboptimal performance. Conversely, the aggregated model demonstrates a test accuracy of $93.87\%$, surpassing the least effective single model by an impressive margin of $58.87\%$.
% This significant improvement underscores the capability of our system to facilitate model owners in  achieving high-quality models.

\begin{figure}[htbp]
	\centering
	\vspace{-0.5em}
	% Subfigure A
	\begin{subfigure}[b]{0.45\linewidth}
		\centering
		\includegraphics[width=0.6\linewidth]{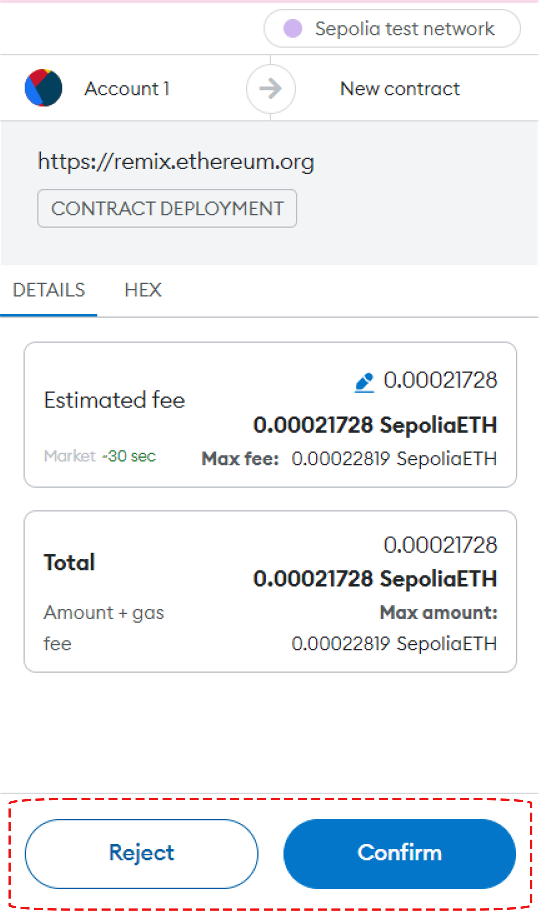}
		\caption{Signing (example)}
		\label{fig:sub-a}
	\end{subfigure}
	%\hfill % This will add a little bit of space between the subfigures horizontally.
	% Subfigure B
	\begin{subfigure}[b]{0.45\linewidth}
		\centering
		\includegraphics[width=0.6\linewidth]{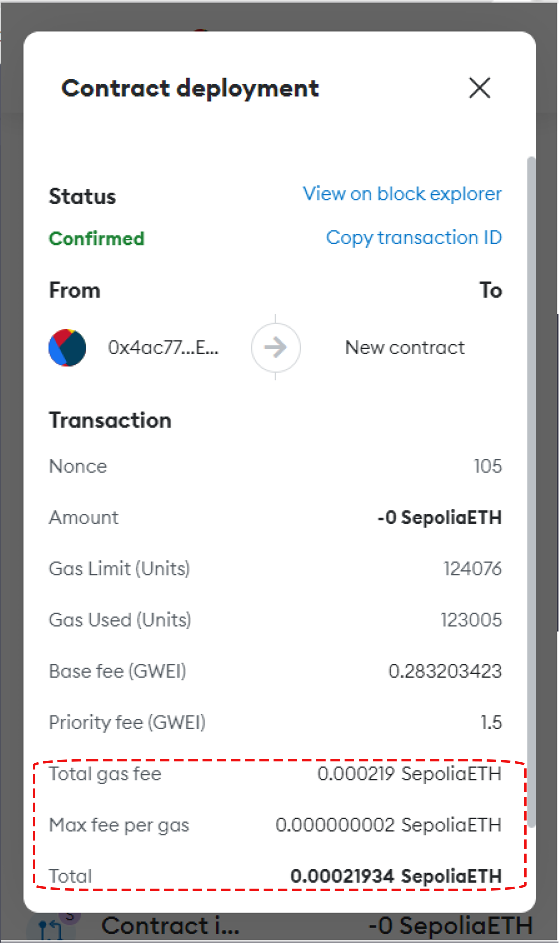}
		\caption{Contract Deployment}
		\label{fig:sub-b}
	\end{subfigure}
	
	\vspace{1em} % This will add some space between the rows of subfigures
	
	% Subfigure C
	\begin{subfigure}[b]{0.45\linewidth}
		\centering
		\includegraphics[width=0.6\linewidth]{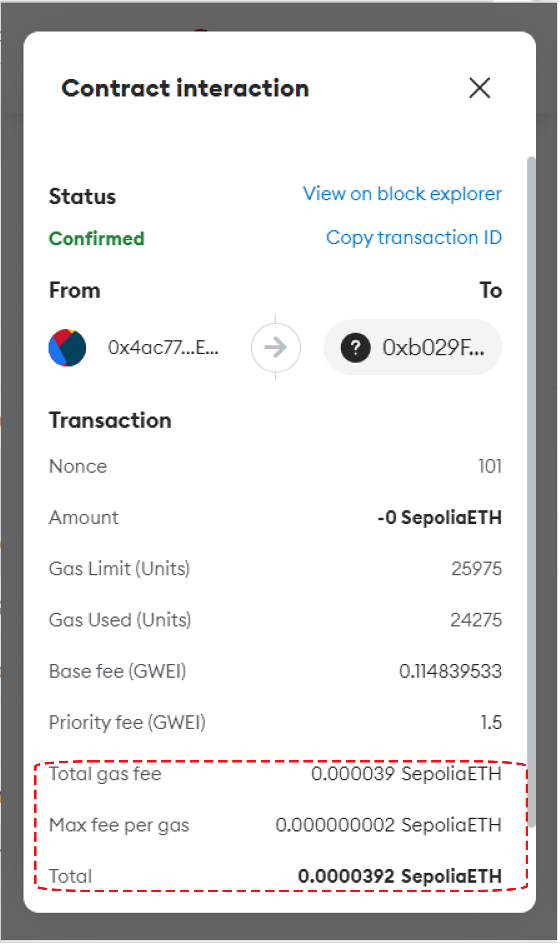}
		\caption{Contract Interaction}
		\label{fig:sub-c}
	\end{subfigure}
	%\hfill
	% Subfigure D
	\begin{subfigure}[b]{0.45\linewidth}
		\centering
		\includegraphics[width=0.6\linewidth]{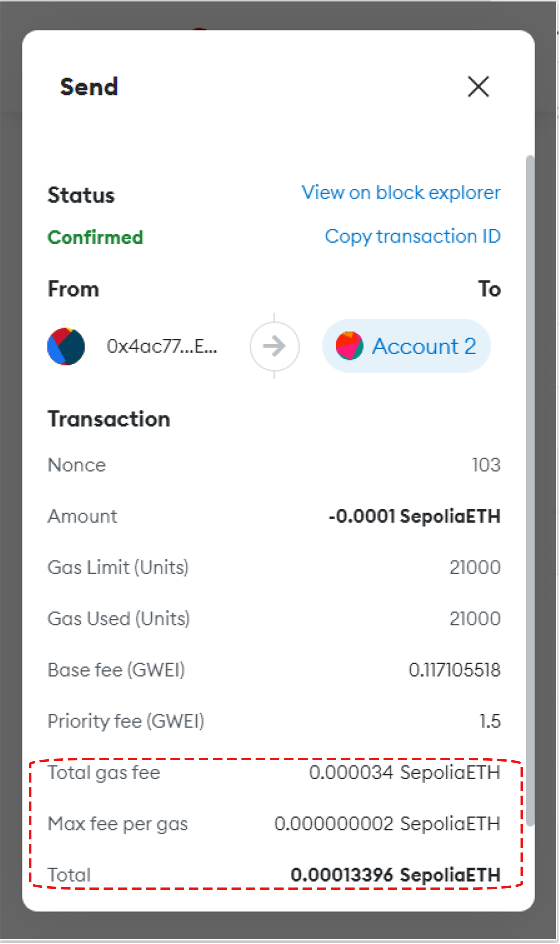}
		\caption{Payment}
		\label{fig:sub-d}
	\end{subfigure}
	\vspace{-1em}
	\caption{The transaction details shown on MetaMask.}
	\label{fig:trans}
		\vspace{-1em}
\end{figure}

\subsection{Transaction Costs}
Our demonstrations outline critical interactions with the smart contract during Steps 1, 4, and 7, each incurring specific gas fees. Model owners deploy the smart contract by clicking the button, which triggers MetaMask to authorize the deployment. Submitting CIDs to the blockchain and transferring ETH to model owners are also facilitated by MetaMask, with model buyers covering the gas fees for these transactions.

Figure~\ref{fig:trans} shows the transaction process via MetaMask, where Figure~\ref{fig:sub-a} details the transaction confirmation phase. Figures~\ref{fig:sub-b}, \ref{fig:sub-c}, and~\ref{fig:sub-d} illustrate the three different transaction types on the blockchain, each with varying total gas fees. From Figures~\ref{fig:sub-b}, \ref{fig:sub-c}, and~\ref{fig:sub-d}, we can see deployment transactions carry the heaviest gas fees (e.g., 0.002 ETH) due to the need to write all functions on the blockchain. For our contract, gas fees for submitting 32-byte CIDs are similar to payment transactions as both involve writing to the blockchain.  Downloading CIDs from the blockchain does not incur gas fees since they don't require data writing.

%Figure~\ref{fig:trans} illustrates the nuances of the transaction details on the MataMask. Figure~\ref{fig:sub-a} captures the confirmation phase, where the transaction is successfully signed once the user clicks the 'confirm' button. Meanwhile, Figure~\ref{fig:sub-b}, \ref{fig:sub-c}, and~\ref{fig:sub-d} display the three distinct types of transactions on the blockchain, highlighting the variability in gas fees among them. The most significant gas fees (i.e., 0.002 ETH) are associated with deployment transactions since they involve writing all functions to the blockchain. However, for our contract, since the model owners are only writing 32-byte CIDs to the blockchain, the incurred gas fees for these contract interactions are comparable to those for payments. It's worth noting that downloading from the blockchain is exempt from gas fees, as it does not involve writing any data to the blockchain. 

\subsection{Payment}

\begin{figure}
	\centering
	\vspace{-1em}
	\includegraphics[width=0.8\linewidth]{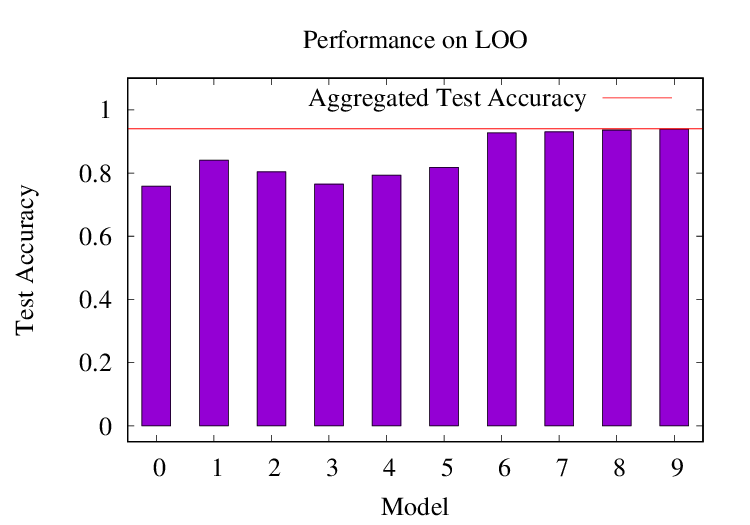}
	\caption{The test accuracy on Leave-one-out (LOO).}
	\label{fig:LOO_accuracy}
	\vspace{-1em}
\end{figure}

After retrieving the models, the model buyers aggregate them and then utilize incentive functions to compute payments. Figure~\ref{fig:LOO_accuracy} shows the test accuracy while any model is dropped. Thus, high test accuracy denoted on model $i$ means less contribution for the model owner $i$. From the figure, we can see that model $7$ is the most useless for the aggregated model. 

\begin{table}[htbp]
	\centering
	\caption{Payment Table}
	\vspace{-1em}
	\label{tab:walletPayments}
	\small
	\begin{tabular}{@{} l r @{}}
		\toprule
		Wallet Address \footnotemark & Payment (ETH) \\ 
		\midrule
		0xbC43368F3062Ba8605A17341d6054CFD649271dD & 0.00162366 \\
		0x5fa7236e596193E2b2cC16b30255C1E7fF9d4957 & 0.00106922 \\
		0x5C892779A6DB3dA3716852Fa2e890B6A9626F159 & 0.00131720 \\
		0x7a305a674Fd11Ad96B56661A6CCe54266f7e2f56 & 0.00157930 \\
		0x0Ea87D03b7C394570000ed84777DeD7468A6Ad48 & 0.00139046 \\
		0xa3Df0eE2026f0448D309Cd8627a8b55Db20e814D & 0.00122177 \\
		0x90341327A3B2Bbe2dDA305d6227d3e3ac6E363D0 & 0.00049194 \\
		0xED0F6C1A47F673A3D087016d48bc1FAf2b557d74 & 0.00046640 \\
		0xeB9865C6FAa7D146C8537005480BeC76d9AF1E03 & 0.00042876 \\
		0x981aDf746f0aF9717CF6f3f42Ad4Cef1b716cEe9 & 0.00041129 \\
		\bottomrule
	\end{tabular}
	\vspace{-1em}
\end{table}

Table~\ref{tab:walletPayments} shows the payment table computed from LOO payment function for 10 model owners. In detail, we allocate the payment based on each participant's contribution, as measured by LOO. 

% In the final step, the model buyer pays 10 model owners with corresponding wallet addresses from the payment table.          

\footnotetext{Note that all wallet addresses are real and can be tracked on the Sepolia Etherscan.  https://sepolia.etherscan.io/.}

\subsection{Overhead Measurement} 
%Now, we measure the computation and communication overheads for whole process. The transmitted information on the blockchain are 32-byte CIDs while in our experiments transmitted model occupies 317Kb on MNIST.

% We measure the total time cost from model owner and model buyer perspective.  For the model owners, the total time cost includes three parts: local model training, uploading the models and sending CIDs to the contract. For the model buyers, the total time cost includes four parts, contract deployment, download CIDs, retrieve models and payment. Note that in the payment, the model buyers computing the payments first and then pay to the model owners.
 
%Figure~xx shows the execution time distribution on model owners and buyers. Note that all model owners are assumed located in the same campus area network.  We can see most of the time cost comes from the interaction on the blockchain. Thus, if we adopt conventional federated learning system with multi-rounds settings, for example, 100 iterations, leading to xxxs overhead. Therefore, one-shot FL is proper for Web 3.0 application. 

We assess the computation and communication overheads of the entire process. Note that on the blockchain, 32-byte CIDs are transmitted, with the models in our experiments occupying 317Kb.

\begin{figure}[htbp]
	\centering
	\vspace{-1em}
	% Subfigure A
	\begin{subfigure}[b]{0.45\linewidth}
	\centering
	\includegraphics[width=1.0\linewidth]{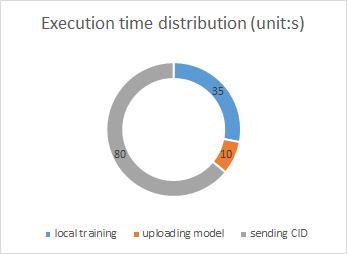}
	\caption{Model Owners}
	\label{fig:time_cost_owner}
	\end{subfigure}
%\hfill % This will add a little bit of space between the subfigures horizontally.
% Subfigure B
	\begin{subfigure}[b]{0.45\linewidth}
	\centering
	\includegraphics[width=1.0\linewidth]{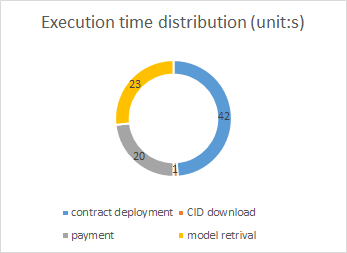}
	\caption{Model Buyers}
	\label{fig:time_cost_buyers}
	\end{subfigure}
	\caption{Execution time distribution on owners and buyers.}
	\label{fig:time_cost}
	\vspace{-1em}
\end{figure}

The total time costs are evaluated from both the model owners' and buyers' perspectives. For model owners, it comprises local model training, model uploading, and sending CIDs to the contract. Model buyers' total time involves contract deployment, CID downloading, model retrieval, and payment processing, where payment calculation precedes the actual transaction.

Figure~\ref{fig:time_cost} presents the time distribution for both model owners and buyers within a unified campus area network, illustrating that the bulk of time consumption is attributed to blockchain interactions. While traditional FL systems may require at least 100 iterations, resulting in significant overhead, our findings endorse that one-shot FL is suitable for Web 3.0 applications.

\begin{comment}
	\begin{figure}
		\centering
		\includegraphics[width=0.8\linewidth]{example-image-a}
		\caption{The fluctuation on gas fee.}
		\label{fig:LOO_accuracy}
	\end{figure} content...
\end{comment}

\section{Conclusion}
In this paper, we introduce OFL-W3, a novel one-shot FL system tailored for Web 3.0 architecture, integrating blockchain technology with smart contracts for efficient transaction management and utilizing the IPFS for decentralized model sharing. Designed to bypass the limitation of the transaction speed and smart contract latency challenges prevalent in existing blockchain frameworks, OFL-W3 demonstrates a viable and innovative approach to implementing one-shot FL in the Web 3.0 context, offering unique insights and potential future directions. Our work not only showcases the practicality of FL applications in a new era of the internet but also sets the stage for further exploration and development within the AI + Web 3.0 domain, promising a transformative impact on both fields.
\begin{comment}
	\begin{acks}
		This work was supported by the [...] Research Fund of [...] (Number [...]). Additional funding was provided by [...] and [...]. We also thank [...] for contributing [...].
	\end{acks}content...
\end{comment}

\begin{acks}
	This research / project is supported by the National Research Foun
	dation, Singapore and Infocomm Media Development Authority under its Trust Tech Funding Initiative, and DSO National Laboratories under the AI Singapore Programme (AISG Award No: AISG2-RP2020-018), and partially supported by the HighTech Support Program from STCSM (No.22511106200), Intel Corporation (UFunding 12679). Any opinions, findings and conclusions or recommendations expressed in this material are those of the author(s) and do not reflect the views of National Research Foundation, Singapore, Infocomm Media Development Authority, STCSM and Intel Corporation. 
\end{acks}
%This research is supported by the National Research Foundation, Singapore and DSO National Laboratories under the AI Singapore Programme (AISG Award No: AISG2-RP2020-018), and partially supported by the HighTech Support Program from STCSM (No.22511106200), Intel Corporation (UFunding 12679). Any opinions, findings and conclusions or recommendations expressed in this material are those of the author(s) and do not reflect the views of National Research Foundation, STCSM and Intel Corporation.

%\clearpage

\bibliographystyle{ACM-Reference-Format}
\bibliography{main}

\end{document}